Transport studies of dual-gated ABC and ABA trilayer graphene: band gap opening and band structure tuning in very large perpendicular electric field

K. Zou[†‡], Fan Zhang[§], C. Clapp[∥], A. H. MacDonald[⊥] and J. Zhu[†#]

[†]Department of Physics, The Pennsylvania State University, University Park, PA 16802

[‡]Department of Applied Physics, Yale University, New Haven, CT 06511

[§]Department of Physics and Astronomy, University of Pennsylvania, Philadelphia, PA 19104

[∥]Department of Chemistry, Amherst College, Amherst, MA 01002

[⊥]Department of Physics, The University of Texas at Austin, Austin, TX 78701

[#]Materials Research Institute, The Pennsylvania State University, University Park, Pennsylvania 16802

**We report on the transport properties of ABC and ABA stacked trilayer graphene using dual, locally gated field effect devices. The high efficiency and large breakdown voltage of the HfO$_2$ top and bottom gates enables independent tuning of the perpendicular electric field and the Fermi level over an unprecedentedly large range. We observe a resistance change of six orders of magnitude in the ABC trilayer, which demonstrates the opening of a band gap. Our data suggest that the gap saturates at a large displacement field of *D* ~ 3 V/nm, in agreement with self-consistent Hartree calculations. In contrast, the ABA trilayer remains metallic even under a large perpendicular electric field. Despite the absence of a band gap, the band structure of the ABA trilayer continues to evolve with increasing *D*. We observe signatures of two-band conduction at large *D* fields. Our self-consistent Hartree calculation reproduces many aspects of the experimental data, but also points to the need for more sophisticated theory.**

Progress in studies of graphene and its few-layer siblings has brought the potential of stacking-structure engineering of two-dimensional materials into sharp focus. For example isolated graphene layers are gapless semiconductors with linear band crossing,[1] whereas Bernal stacked bilayer graphene has hyperbolic bands[1-4] with an electric-field tunable band gap.[5,6] Trilayer graphene possesses two stable stacking orders: ABA (Bernal) and ABC (rhombohedral) stacking, which have distinct electronic properties that are of interest both for fundamental science and for technological applications.[7-23] The conduction and valence band dispersions of ABC-trilayer graphene (ABC-TG) are expected to have approximately cubic dispersion and to touch at the charge neutrality point (CNP).[8,12] The associated divergent density of states leads to an interaction-driven broken symmetry ground state.[13] In the presence of a perpendicular magnetic field, the approximate 12-fold degeneracy of the $0^{th}$ Landau level is expected to be lifted by interactions.[18,19,21] From the technological standpoint, ABC-TG is predicted to have an electric field tunable band gap up to 0.3 eV,[10,12] which is attractive for optoelectronic applications such as infrared photodetectors. Demonstrating control of the band gap and understanding the transport properties of gapped ABC-TG are critical to realize these applications. Although a tunable band gap in ABC-TG has been demonstrated by infrared absorption spectroscopy,[16] transport characteristics are less well established and limited to small electric field.[14,22] In stark contrast to ABC-TG, the energy dispersion of ABA-trilayer graphene (ABA-TG) consists of both monolayer-like and bilayer-like subbands.[9] Interestingly, both subbands become massive due to remote hopping processes allowed by mirror symmetry with respect to the middle layer,[9] though ABA-TG remains gapless. Most studies to date have focused on ABA-TG's unique Landau level structures in a magnetic field.[17,20,21,23] ABA-TG has been shown to remain metallic upon the application of a perpendicular electric field but little is known beyond that.[7,14,20] Multiple bands contribute to charge transport in ABA-TG even close to the charge neutrality point. The characteristics of these bands, and in particular their evolution in a perpendicular electric field, are still poorly understood.

In this work, we report on the transport properties of ABC-TG and ABA-TG using dual locally gated field effect devices. By controlling the displacement field $D$ and the carrier density $n$ independently over a wide range using the two gates, we demonstrate unambiguously the opening of a band gap in ABC-TG, where the resistance at the CNP increases by five orders of magnitude with increasing $D$. Furthermore, our measurement suggests a saturation of the band gap at $D \sim 3.0$ V/nm, in agreement with a tight-binding calculation that takes Hartree screening into account self-consistently. The signatures of two-band conduction are observed for the first time in ABA-TG. The onset carrier density as a function of $D$ is compared to a self-consistent Hartree theory. The agreement between experiment and theory is satisfactory; we argue that the discrepancy is likely due to an inadequate treatment in the simplified calculations of the occupied sigma bands and of exchange interactions.

The procedures used to fabricate the dual-gated devices are described in the Supporting Information. Briefly, we first pattern local bottom gate stacks of 28 nm $HfO_2$ film and 30 nm Au on $SiO_2$/doped Si substrate using optical lithography, reactive ion etching, metal deposition and atomic layer deposition of $HfO_2$. Graphene sheets are then mechanically exfoliated to the local gate area using bulk graphite (ZYA grade, distributed by SPI). A TG piece is first identified by its optical contrast and confirmed by Raman spectroscopy using the profile of its 2D band[15] (Figure 1b). Hall bar devices are made by e-beam lithography. After the deposition of metal electrodes, another 28 nm thick $HfO_2$ film is deposited on top of the device as the top gate dielectric, followed by the patterning and deposition of local top gate electrodes. Both $HfO_2$ layers are grown by atomic layer deposition using the recipe previously described in ref 24. An optical micrograph of a finished device is shown in Figure 1a, together with a schematic of the device as the inset. Micro Raman spectra along the length of the device reveals that it consists of ABA-TG and ABC-TG regions separated by a domain boundary under top gate 2. The two corresponding 2D band profiles are shown in Figure 1b. This identification is further supported by resistance measurements using different pairs of voltage probes (Figure 1a) and described in detail later. This unique geometry enables us to directly compare the transport properties of TG samples with different stacking orders on the same device.

Both the top and bottom HfO$_2$ gates exhibit a gating efficiency of ~ 5.5x10$^{12}$/cm$^2$ per Volt, as deduced from the sample's quantum Hall sequence at a fixed magnetic field. This is approximately 80 times higher than the commonly used SiO$_2$ global back gate. The breakdown voltages of both gates are greater than 8 V. The large capacitance and high breakdown voltage of the two gates allow us to vary the perpendicular displacement field $D$ and the carrier density $n$ independently over exceedingly large range with $|D|$ up to 6 V/nm and $n$ up to 3.3 x10$^{13}$/cm$^2$ in experiments reported here, thereby probing regimes unexplored in previous studies limited to $|D| \leq 1$ V/nm.[7, 13, 14, 17, 18, 20, 22] We measure the resistance $R$ between different pairs of voltage probes in the temperature range of 1.5 K to 200 K while varying the top and bottom gate voltage $V_{tg}$ and $V_{bg}$. Both four-terminal standard lock-in techniques and two-terminal DC techniques are used to measure the resistances as they vary by six orders of magnitude in different portion of the device. We estimate the carrier mobility to be about 1000 cm$^2$/Vs. The technical details are given in the Supporting Information.

Figure 2a plots the resistance between probes 2 and 3, $R_{23}$ ($V_{tg}$) at $T$ = 1.5 K, as a function of sweeping $V_{tg}$ from -5.2 V to 5.8 V on top gate 2 and at fixed $V_{bg}$'s ranging from -4.0 V to 5.0 V in 0.5 V steps. The respective displacement field of the top and bottom gate (+ for top and – for bottom) $D_{tg,bg} = \pm\varepsilon(V_{tg,bg} - V^0_{tg,bg})/d$, where the HfO$_2$ film thickness $d$ = 28 nm is obtained with atomic force microscopy and its dielectric constant $\varepsilon$ = 28 is calculated using the charging efficiency of the gates. $V^0_{tg}$ = 1.0 V and $V^0_{bg}$ = 0.5 V are due to the unintentional chemical doping from the environment. Similar to bilayer graphene, the maximum of each trace in Figure 2a corresponds to a charge neutrality point (CNP), where $n \sim (D_{tg} - D_{bg}) = 0$. As the average perpendicular field through the trilayer $D = (D_{tg} + D_{bg})/2$ increases from 0 to $\pm$ 4.5 V/nm, $R_{23}$ at the CNP increases dramatically by five orders of magnitude from 6000 Ω to several hundred MΩ. Although $R_{23}$ consists of both ABA-TG and ABC-TG, the dramatic gate dependence originates from that of the ABC portion (see Figure 3 and Supporting Information for more discussion). Away from the CNP, $R_{23}$ sharply decreases to a few hundred Ω as the carrier density *n* increases. The ratio between the resistance of the "on" state at high carrier density and the "off" state at the CNP reaches $10^6$ at large $|D|$, which is the largest ratio reported for trilayer graphene. The observed gate dependence of $R_{23}$ unambiguously demonstrates the opening of a band gap in the ABC-TG portion of the device.

Figure 2a also reveals an interesting trend of the CNP resistance with increasing $|D|$. As the dashed lines in Figure 2a show, the CNP resistance follows an approximate exponential dependence on $|D|$ at small fields but deviates at $|D| > 3.0$ V/nm. This trend is more clearly seen in Figure 2b, where we plot $R_{23}^{CNP}$ vs. $|D|$ in a semi-log plot (left axis). Also plotted there (right axis) is the calculated band gap $E_g$ in ABC-TG using a density functional theory.[12] For $|D| > 1$ V/nm, where studies of bilayer graphene suggest [25] that $E_g$ is likely to be no longer dominated by disorder, $\ln(R_{23}^{CNP})$ and $E_g$ have similar dependences on $|D|$. Both increase roughly linearly with $|D|$ until $|D|$ reaches approximately 3 V/nm, where a tendency toward saturation is seen both in theory and in experiment. Although a quantitative relation between $R_{23}^{CNP}$ and $E_g$ is difficult to establish due to the complicating role of disorder, the saturation of $R_{23}^{CNP}$ at large l$D$l is consistent with the saturation of the band gap. In the Supporting Information, we demonstrate the saturation of $R_{23}^{CNP}$ is not due to parasitic resistance or heating effects and other relevant energy scales related to $E_g$ also saturate with increasing $|D|$. Together, these are the first experimental indications of the saturation of $E_g$ in ABC-TG.

The dependence of the transport properties of the ABA portion of our device on *D* is drastically different. Figure 3 plots the resistance between probes 1 and 2, $R_{12}(V_{tg})$ at *T* = 1.5 K as a function of $V_{bg}$ and $V_{tg}$ on top gate 1, taken within the same voltage range and interval as in Figure 2a. $R_{12}$ measures ABA-TG only. In stark contrast to the drastic increase of $R_{23}^{CNP}$ in ABC-TG with $|D|$, the maxima of $R_{12}(V_{tg})$, $R_{12}^{max}$ is several kΩ. This observation unambiguously demonstrates the lack of an electric field-induced band gap in ABA-TG, in agreement with previous experiments[7, 14, 20] and the theoretical expectation[9] of a semi-metallic band structure for this stacking order.

Figure 3 also reveals several interesting features not seen in previous studies of dual-gated ABA-TG.[7, 14, 20] $R_{12}^{max}$ of each curve (marked by blue triangles and purple squares) first increases with increasing |D|, although this increase is much smaller than that of the ABC-TG over the same range, then reaches a global maximum near |D| = 2 V/nm, and eventually decreases with further increase of |D|. Furthermore, in conjunction with the decrease of $R_{12}^{max}$ at large |D|, a "side" peak starts to develop. This new peak, marked by red triangles in Figure 3, occurs for both negative and positive D's and appears always to the right side of $R_{12}^{max}$.

To elucidate the origins of the two peaks, we plot in Figure 4a $V_{tg}$ and $V_{bg}$ of the marked positions in Figure 3 (red triangles, blue triangles and purple squares). Surprisingly, the "side" peaks marked by the red triangles follow a straight line of slope one (the dashed line in Figure 4a), which also coincides with the $V_{tg}$ dash $V_{bg}$ line of the CNPs on the ABC portion of the device. In another word, the "side" peaks in Figure 3 *actually* correspond to the CNPs of the ABA-TG, contrary to the intuition of associating $R_{12}^{max}$ with the CNP. The position of $R_{12}^{max}$ (blue triangles) itself increasingly deviates from the CNP line as |D| increases. At |D| < 2.5 V/nm, only one resistance maximum is perceptible, as marked by the purple triangles in Figure 4a. Evidently, the non-monotonic dependence of $R_{12}^{max}$ on D originates from the evolution of the two peaks.

The characteristics of $R_{12}$ ($V_{tg}$) at large |D| are reminiscent of two-band conduction observed in GaAs/GaAlAs heterostructures[26] and more recently in bilayer graphene.[27] In a 2D system with more than one subband, as $E_F$ approaches the band edge of the higher subband, additional scattering channels become available, causing resistance to increase. As $E_F$ further increases, carriers occupy the higher subband and participate in both screening and conduction; these effects cause the total resistance to drop again. This process results in the "negative differential resistance" region in the transfer curve plotted in Figure 3.

A similar two-band conduction model that takes into account the evolution of ABA-TG bands in a perpendicular D field provides a natural explanation for the appearance and evolution of the two resistance peaks observed in Figure 3. This situation is illustrated in Figures 4b and 4c, where we plot the four low energy bands of ABA-TG for $U_{SCR}$ = 0 and $U_{SCR}$ = 160 meV respectively, using a tight-binding model with the hopping parameters as follows:[17] $\gamma_0$ = 3 eV, $\gamma_1$ = 0.4 eV, $\gamma_2$ = -0.028 eV, $\gamma_3$ = $\gamma_4$ = 0, $\gamma_5$ = 0.05 eV and $\delta$ = 0.046 eV. Here, $U_{SCR}$ is the screened potential difference between the top and bottom layers of the ABA-TG. As $D$ increases, so do $U_{SCR}$ and the splitting of the higher energy subbands as shown in Figures 4b and 4c. Starting from the CNP (dashed lines in Figures 4b and 4c), $E_F$ moves downward as $V_{tg}$ becomes more negative and more holes are added. Consequently, $R_{12}$ decreases. As $E_F$ approaches the 2nd valance band, increased scattering causes $R_{12}$ to turn around and increase until the 2nd valance band is populated and the added conduction channel causes $R_{12}$ to drop again. Thus, we associate the blue triangles with $E_F$ reaching the top of the 2nd valence band, as illustrated by the solid line in Figure 4c using the corresponding experimental curve. The width of the negative differential resistance (NDR) region in Figure 3 is controlled by charge disorder, which produces localized states near the band edges and also smears their positions.[28, 29] This disorder broadening of subbands also causes the merge of the two resistance peaks at small field $|D|$ < 2.2 V/nm (Figure 4b), where the band splitting is too small to be resolved.[25, 28-30] Interestingly, we did not see evidence of NDR occurring at the bottom of the 2nd conduction band. This is likely due to the much smaller density of states (DoS) available for scattering, since this band evolves from a monolayer-like band while the 2nd valance band from a bilayer-like band.

To quantitatively compare our experiment with the commonly used self-consistent Hartree theories, we deduce the screened $U_{SCR}$ in the tight-binding model[11, 12] that yields the population of the 2nd valance band at the carrier density measured at $R_{12}^{max}$. The resulting $U_{SCR}$ (blue symbols) and the $E_F$ (red symbols) that corresponds to the top of the 2nd valence band are plotted in Figure 4d as a function of $|D|$ at $R_{12}^{max}$ (bottom axis), and the unscreened external potential difference $U_{ext}$ = $2de|D|$ (top axis), where $d$ = 0.335 nm is the interlayer spacing of trilayer. Also plotted there are $U_{SCR}$ (blue solid line) and $E_F$ (red solid line) obtained in our self-consistent Hartree theory with the tight-binding parameter values cited earlier. Here the calculations use self-consistent Hartree screening to solve for $U_{SCR}$ and $E_F$ required to populate the 2nd valence band at the $U_{ext}$ given by experiment. Screening by the remote sigma bands not included in our calculation is modeled by a fixed dielectric constant $\varepsilon$ = 4.

Overall, the theory results qualitatively reproduce the experimental trend of screening very well, although the calculated values of $U_{SCR}$ and $E_F$ are larger by ~ 30%. Varying the tight binding parameters will change both the calculated and experimentally determined $U_{SCR}$, but in similar fashions that will not reduce the discrepancy. Experimentally, disorder broadening hinders precise determination of the band edges. This uncertainty does not appear to be large enough to explain the discrepancy. Theoretically, the overestimate of $U_{SCR}$ might be attributed to an underestimate of screening. This could originate from the inadequate treatment of the sigma band, whose role in polarization is more complex and could be stronger than our phenomenological $\varepsilon = 4$ effect.[16, 31] Our neglect of exchange energies may also have an effect. Our experiment provides the first data set in ABA-TG to which theoretical calculations may be compared.

In summary, we report on the electrical transport of dual-gated ABA and ABC trilayer graphene in exceedingly large perpendicular electric field. Our experiments unambiguously demonstrate the opening of a large transport band gap in ABC-TG, where the resistance changes by six orders of magnitude. Our data suggest the band gap saturates at $|D|$ ~ 3 V/nm. In ABA-TG, we observe signatures of two-band conduction at large $D$. Tight-binding calculations using self-consistent Hartree screening can explain the observations qualitatively, but reveal quantitative discrepancies which motivate further theoretical work.

We are grateful for helpful discussions with Qiuzi Li. We thank Bei Wang for the Raman measurements. K.Z. and J.Z. are supported by ONR under grant No. N00014-11-1-0730. F. Z. is supported by DARPA under grant SPAWAR N66001-11-1-4110. C. C. is supported by an NNIN REU grant. A.H.M. is supported by Welch Foundation grant TBF1473 and DOE Division of Materials Sciences and Engineering grant DE-FG03-02ER45958. The authors acknowledge use of facilities at the PSU site of NSF NNIN.

1.      Castro Neto, A. H.; Guinea, F.; Peres, N. M. R.; Novoselov, K. S.; Geim, A. K. *Reviews of Modern Physics* **2009,** 81, (1), 109-162.

2. Henriksen, E. A.; Eisenstein, J. P. *Physical Review B* **2010,** 82, (4), 041412.
3. Zou, K.; Hong, X.; Zhu, J. *Physical Review B* **2011,** 84, (8), 085408.
4. Young, A. F.; Dean, C. R.; Meric, I.; Sorgenfrei, S.; Ren, H.; Watanabe, K.; Taniguchi, T.; Hone, J.; Shepard, K. L.; Kim, P. *Physical Review B* **2012,** 85, (23), 235458.
5. McCann, E.; Fal'ko, V. I. *Physical Review Letters* **2006,** 96, (8), 086805.
6. Zhang, Y. B.; Tang, T. T.; Girit, C.; Hao, Z.; Martin, M. C.; Zettl, A.; Crommie, M. F.; Shen, Y. R.; Wang, F. *Nature* **2009,** 459, (7248), 820-823.
7. Craciun, M. F.; RussoS; YamamotoM; Oostinga, J. B.; Morpurgo, A. F.; TaruchaS. *Nat Nano* **2009,** 4, (6), 383-388.
8. Koshino, M.; McCann, E. *Physical Review B* **2009,** 80, (16), 165409.
9. Koshino, M.; McCann, E. *Physical Review B* **2009,** 79, (12), 125443.
10. Avetisyan, A. A.; Partoens, B.; Peeters, F. M. *Physical Review B* **2010,** 81, (11), 115432.
11. Koshino, M. *Physical Review B* **2010,** 81, (12), 125304.
12. Zhang, F.; Sahu, B.; Min, H.; MacDonald, A. H. *Physical Review B* **2010,** 82, (3), 035409.
13. Bao, W.; Jing, L.; Velasco, J.; Lee, Y.; Liu, G.; Tran, D.; Standley, B.; Aykol, M.; Cronin, S. B.; Smirnov, D.; Koshino, M.; McCann, E.; Bockrath, M.; Lau, C. N. *Nature Physics* **2011,** 7, (12), 948-952.
14. Jhang, S. H.; Craciun, M. F.; Schmidmeier, S.; Tokumitsu, S.; Russo, S.; Yamamoto, M.; Skourski, Y.; Wosnitza, J.; Tarucha, S.; Eroms, J.; Strunk, C. *Physical Review B* **2011,** 84, (16), 161408.
15. Lui, C. H.; Li, Z. Q.; Chen, Z. Y.; Klimov, P. V.; Brus, L. E.; Heinz, T. F. *Nano Letters* **2011,** 11, (1), 164-169.
16. Lui, C. H.; Li, Z. Q.; Mak, K. F.; Cappelluti, E.; Heinz, T. F. *Nature Physics* **2011,** 7, (12), 944-947.
17. Taychatanapat, T.; Watanabe, K.; Taniguchi, T.; Jarillo-Herrero, P. *Nature Physics* **2011,** 7, (8), 621-625.
18. Zhang, F.; Jung, J.; Fiete, G. A.; Niu, Q.; MacDonald, A. H. *Physical Review Letters* **2011,** 106, (15), 156801.
19. Zhang, L. Y.; Zhang, Y.; Camacho, J.; Khodas, M.; Zaliznyak, I. *Nature Physics* **2011,** 7, (12), 953-957.
20. Henriksen, E. A.; Nandi, D.; Eisenstein, J. P. *Physical Review X* **2012,** 2, (1), 011004.
21. Zhang, F.; Tilahun, D.; MacDonald, A. H. *Physical Review B* **2012,** 85, (16), 165139.
22. Khodkov, T.; Withers, F.; Hudson, D. C.; Craciun, M. F.; Russo, S. *Appl. Phys. Lett.* **2012,** 100, (1), 013114.
23. Lee, Y.; Velasco, J.; Tran, D.; Zhang, F.; Bao, W.; Jing, L.; Myhro, K.; Smirnov, D.; Lau, C. N. *arXiv:1210.6592* **2012**.
24. Zou, K.; Hong, X.; Keefer, D.; Zhu, J. *Physical Review Letters* **2010,** 105, (12), 126601.
25. Zou, K.; Zhu, J. *Physical Review B* **2010,** 82, (8), 081407.
26. Stormer, H. L.; Gossard, A. C.; Wiegmann, W. *Solid State Communications* **1982,** 41, (10), 707-709.
27. Efetov, D. K.; Maher, P.; Glinskis, S.; Kim, P. *Physical Review B* **2011,** 84, (16), 161412.
28. Nilsson, J.; Castro Neto, A. H. *Physical Review Letters* **2007,** 98, (12), 126801.
29. Rossi, E.; Das Sarma, S. *Physical Review Letters* **2011,** 107, (15), 155502.
30. Li, Q.; Hwang, E. H.; Das Sarma, S. *Physical Review B* **2011,** 84, (11), 115442.
31. Siegel, D. A.; Park, C. H.; Hwang, C.; Deslippe, J.; Fedorov, A. V.; Louie, S. G.; Lanzara, A. *Proceedings of the National Academy of Sciences of the United States of America* **2011,** 108, (28), 11365-11369.

Figure 1: Device layout and Raman Spectra. (a) An optical micrograph of the dual gated trilayer graphene device. The blue shaded area represents the trilayer flake. The six voltage probes and two top gate electrodes are marked on the graph. The local bottom gate is not visible in this image. The dashed line marks a schematic location of the domain boundary between the ABA and ABC stacked portion of the flake. Inset: a schematic of the cross section. (b) Raman spectra of the 2D band taken at the black and red spots marked in (a) respectively. The black trace shows the profile of an ABA-TG and the red trace shows that of an ABC-TG.

Figure 2: Transport characteristics of the ABC-TG. (a) Semi-log plot of $R_{23}$ as a function of $V_{bg}$ and $V_{tg}$. From left to right, $V_{bg}$ changes from 5.0 V to - 4.0 V in 0.5 V steps. The top axis marks the average $D$ at the CNPs. The dashed lines are guide to the eye. $T$ = 1.5K. (b) Left axis: Semi-log plot of the CNP resistance $R_{23}^{CNP}$ as a function of $|D|$. Black squares are from positive $D$'s and red circles are from negative $D$'s. Right axis: Theoretical band gap $E_g(D)$ of ABC-TG from ref 12.

Figure 3: Transport characteristics of the ABA-TG. $R_{12}$ as a function of $V_{bg}$ and $V_{tg}$. From left to right, $V_{bg}$ changes from 5.0 V to - 4.0 V in 0.5 V steps. $T$ = 1.5 K. The red triangles mark the CNPs. The blue triangles mark the top of the $2^{nd}$ valence band. The purple squares mark the merged peak. See the main text for explanations.

Figure 4: Multiband conduction in ABA-TG. (a) Positions of $R_{12}$ peaks marked in Figure 3 in a $V_{tg}$ vs. $V_{bg}$ plot. The dashed line corresponds to slope 1 and coincides with the CNP line of the ABC portion of the device. (b) and (c) The low-energy bands of ABA-TG obtained using a tight-binding model with $U_{SCR}$ = 0 meV (b) and $U_{SCR}$ = 160 meV (c). $R_{12}(V_{tg})$ data are plotted on the side to qualitatively illustrate the resistance change with changing $E_F$. The symbols follow Figure 3. The gray dashed lines in (b) and (c) mark the CNP point and the gray solid line in (c) marks the edge of the $2^{nd}$ valence band. (d) $U_{SCR}$ (left axis) and $E_F$ (right axis) as a function of $|D|$ (bottom axis) and $U_{ext}$ (top axis) obtained by fitting experimental data (symbols) and from self-consistent Hartree screening calculations (lines). Solid squares are from positive $D$'s and hollow squares are from negative $D$'s.

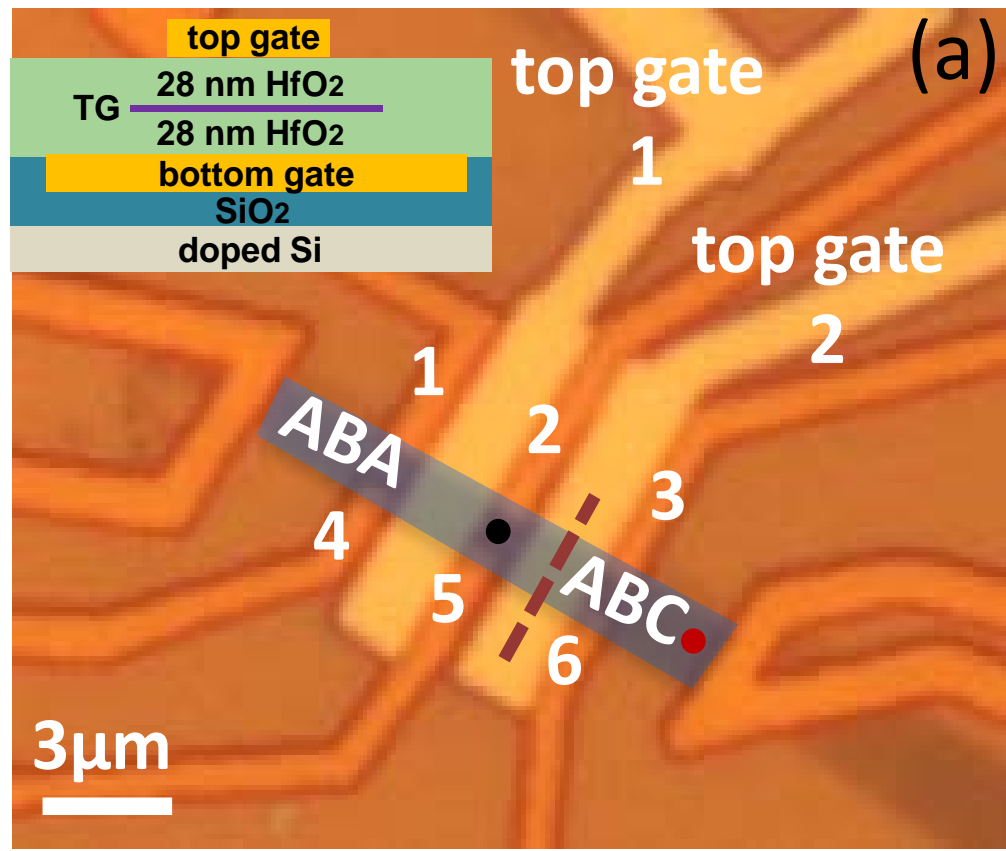
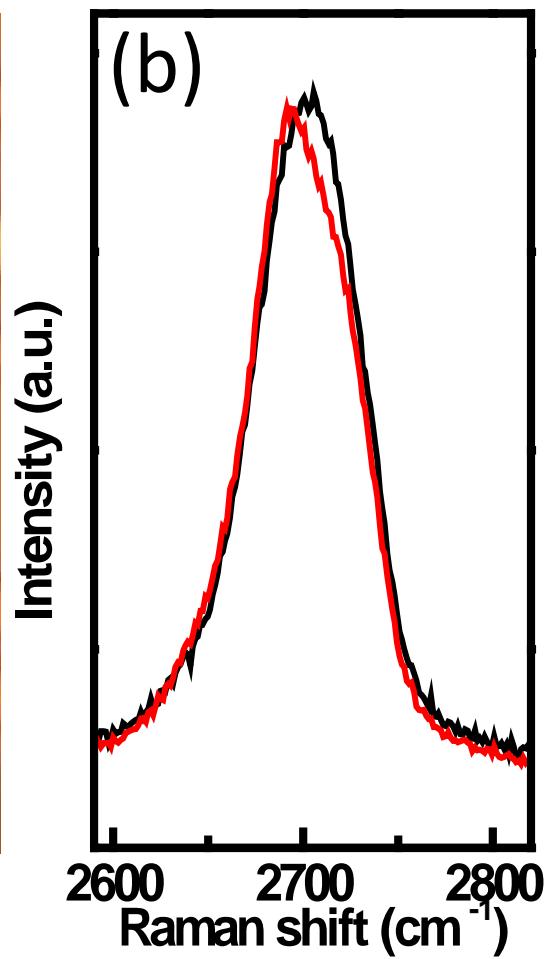

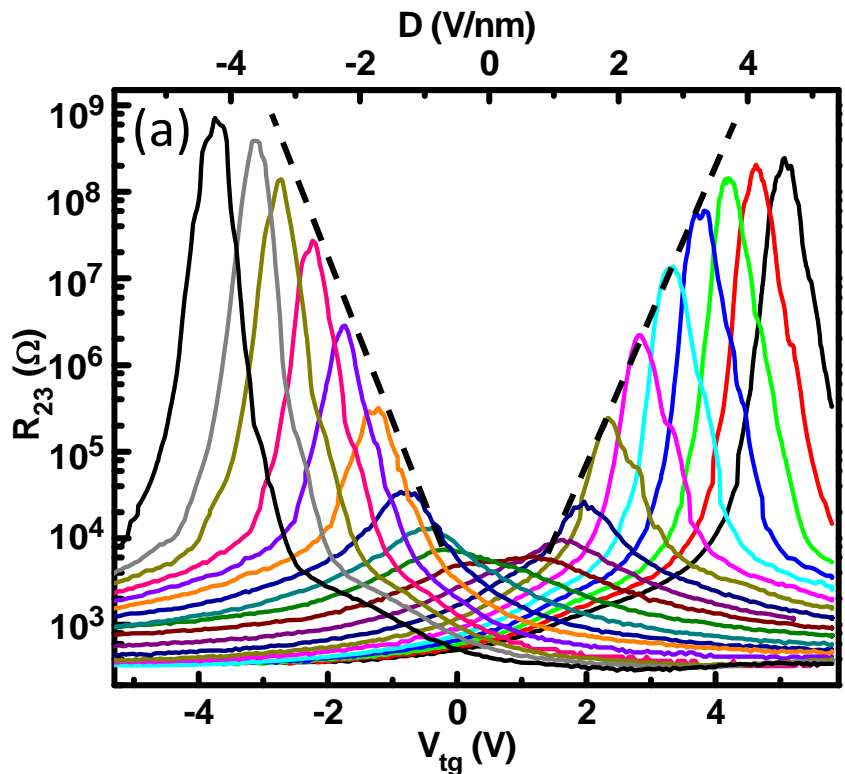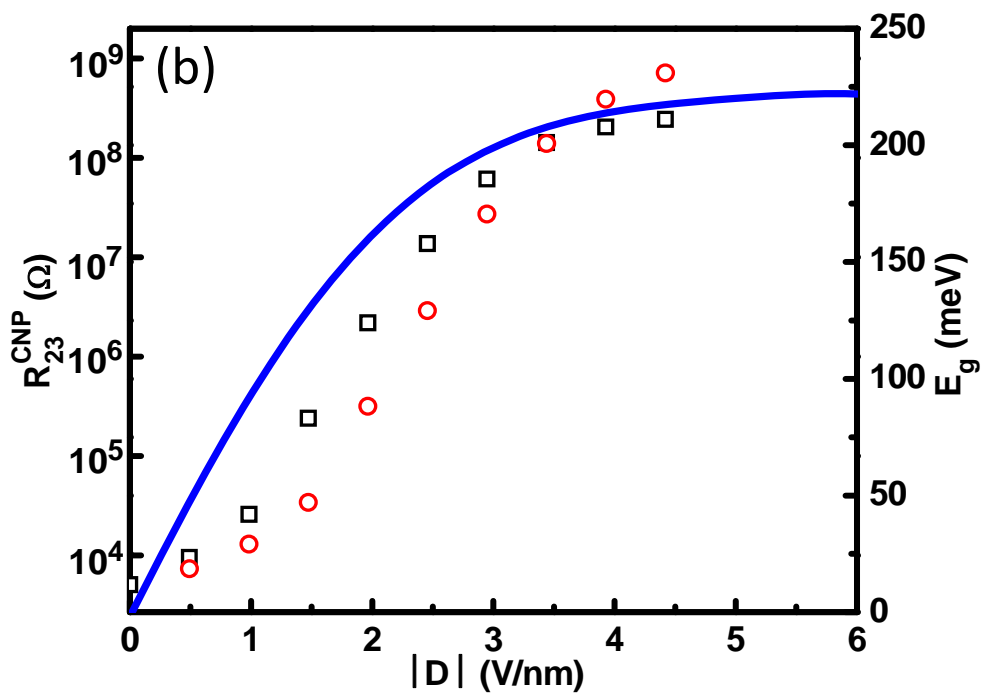

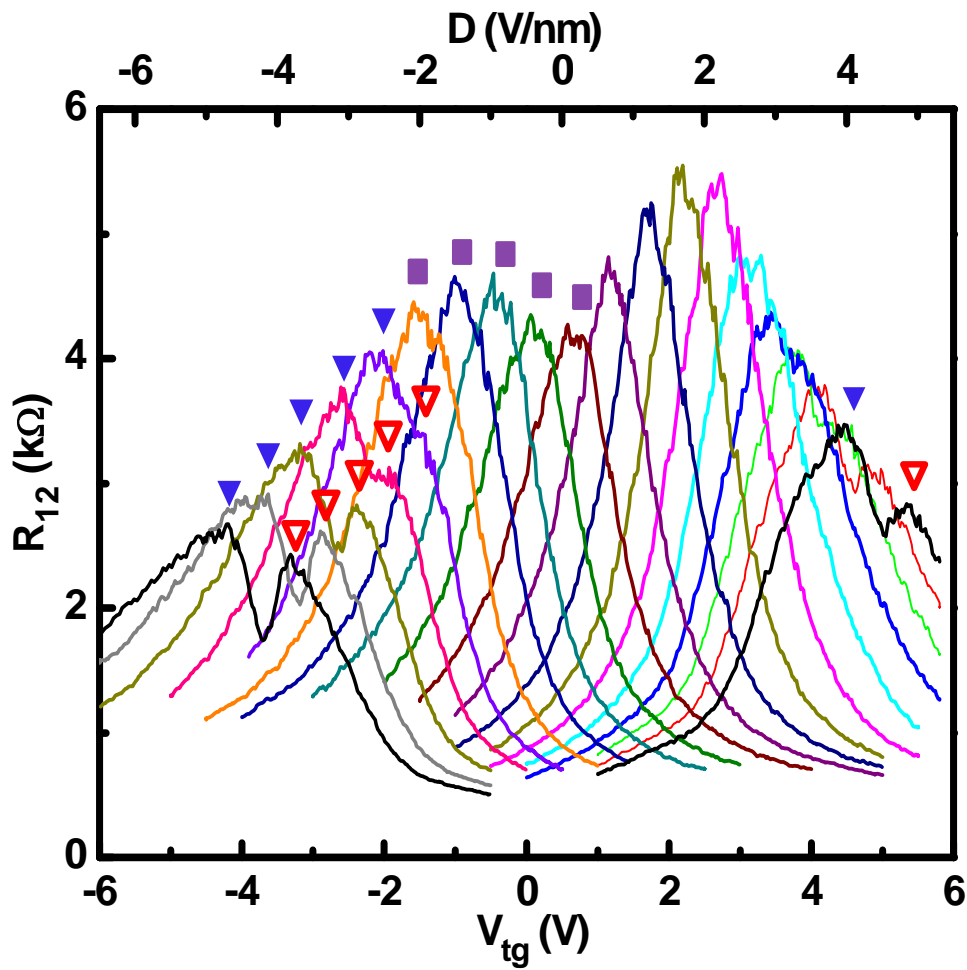

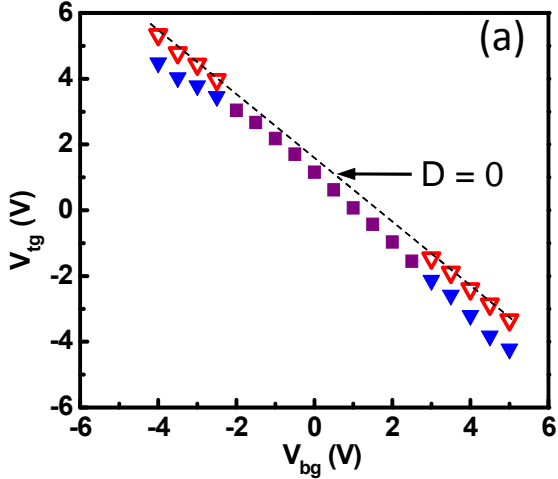
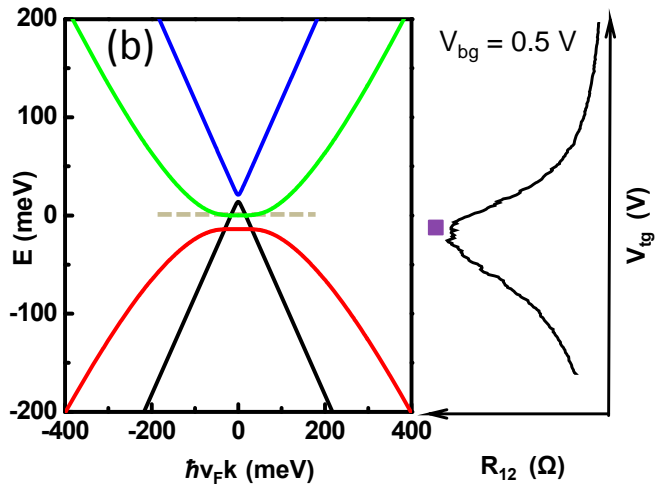
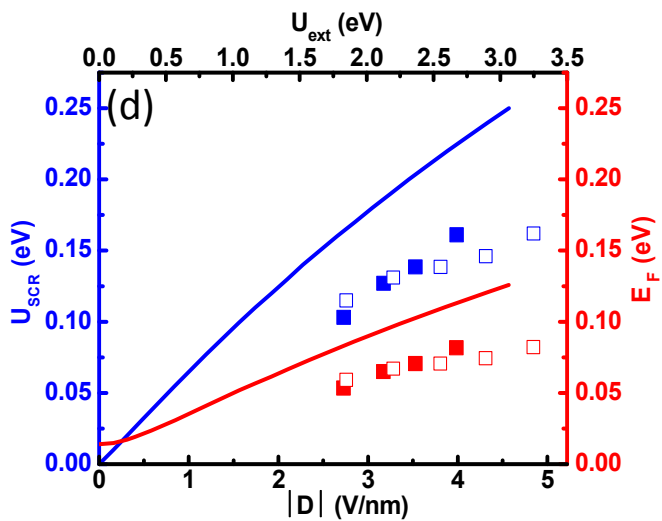
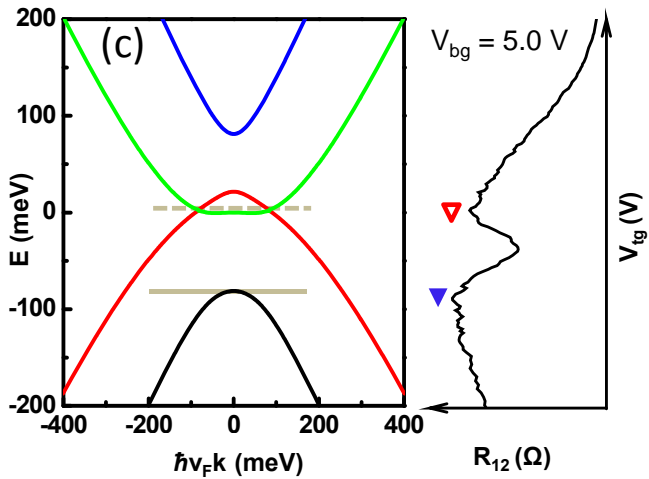

# Transport studies of dual-gated ABC and ABA trilayer graphene: band gap opening and band structure tuning in very large perpendicular electric field

## Supporting Information

K. Zou[†‡], Fan Zhang[§], C. Clapp[∥], A. H. MacDonald[⊥] and J. Zhu[†#]

[†]Department of Physics, The Pennsylvania State University, University Park, PA 16802

[‡]Department of Applied Physics, Yale University, New Haven, CT 06511

[§]Department of Physics and Astronomy, University of Pennsylvania, Philadelphia, PA 19104

[∥]Department of Chemistry, Amherst College, Amherst, MA 01002

[⊥]Department of Physics, The University of Texas at Austin, Austin, TX 78701

[#]*Materials Research Institute, The Pennsylvania State University, University Park, Pennsylvania 16802*

1. Device fabrication procedure.

The local bottom gate electrodes are first defined by optical lithography on standard 290 nm $SiO_2$/doped Si wafers. We then use reactive ion etch ($Cl_2$ 20 sccm, 20 mTorr, inductive power 50 W, bias power 200 W) to etch a trench of approximately 40 nm in depth. 40 nm Ti/Au is then deposited into the trench using physical vapor deposition, followed by lift-off. The wafer surface is cleaned by UV/ozone and rinsed in acetone and IPA sequentially to remove residue from the lithography. We then grow 28 nm of $HfO_2$ using ALD recipes described in ref 1. Graphene flakes are then mechanically exfoliated to the $HfO_2$ surface using bulk graphite (ZYA grade, distributed by SPI). Pieces landing above the bottom gate electrodes are screened optically and by Raman spectroscopy for the appropriate size and number of layers. From this point on, we follow procedures described in ref 1 to pattern the electrodes to the sample, grow another 28 nm of $HfO_2$ and pattern the top gate electrode.

2. Measurement techniques.

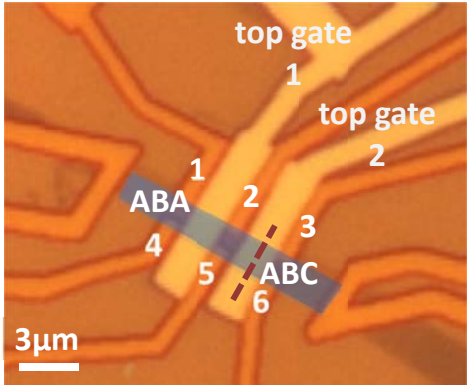

Figure S1. This is the same micrograph given in Figure 1 of the main text. Note the location of the domain boundary between the ABA and ABC stacked TG relative to the six voltage probes and the two top gates marked in the figure.

Four-terminal standard AC lock-in techniques (47 Hz) are used to measure the resistance $R_{12}$, which only includes ABA-TG.

The area between probes 2 and 3 contains both ABA and ABC stacked TG. Because of the drastic gate dependence the ABC-TG displays, $R_{23}$ ($V_{tg}$) shown in Figure 2a combines four-terminal lock-in measurements for resistance below 100 kΩ and two-terminal DC resistance measurements for resistance above 100 kΩ. In the AC measurements, the excitation current is adjusted as $R_{23}$ changes to avoid Joule heating (0.2 – 50 nA). In the DC measurements, we source voltage ($V$) between the two current probes and measure the current ($I$) through the whole flake. Figure S2 shows an example of such $I(V)$ data. Both positive and negative biases are averaged to remove a small offset near the zero bias. The $I(V)$ is linear up to ~ 4 mV. Measurements of $R_{23}$ at higher temperatures are obtained using $V = \pm 0.5$ mV, which is well within the linear regime of the $I(V)$.

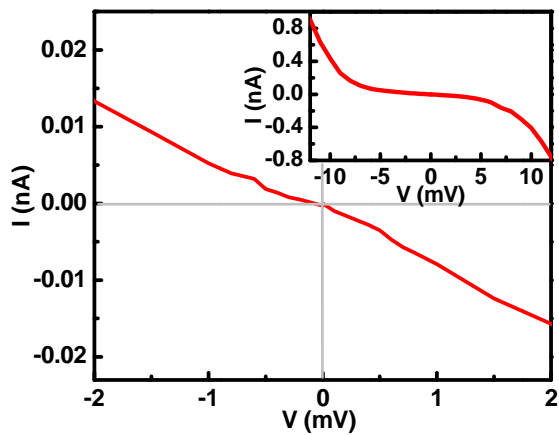

Figure S2. $I$($V$) curve at $D \sim 4.5$ V/nm. $T$ = 1.5K. Inset: The same curve on a larger scale including the non-linear regime.

Using $R_{12}$ data shown in Figure 3, we estimate that the resistance of the whole ABA-TG portion of the flake is always less than 15 kΩ, with the resistance of the area underneath top gate 2 no more than 5 kΩ. The resistance of the ABC-TG sheet not underneath top gate 2 also remains less than 5 kΩ because it's heavily doping by $V_{bg}$. Thus, at large $D$ field, the DC resistance is dominated by the dual-gated ABC-TG underneath top gate 2. Indeed, both four-terminal and two-terminal measurements yield nearly identical values in the vicinity of 100 kΩ, above which data in Figure 2a reflect the gate dependence of the dual-gated ABC-TG only. In the 10 kΩ range, $R_{23}$ includes contributions from both the ABA-TG and the ABC-TG and care must be taken in analyzing data in this regime.

3. Additional evidence of band gap saturation in ABC-TG.

In Figure 2b of the text, we show that $R_{23}^{\mathrm{CNP}}(D)$ tends toward saturation with increasing $D$. Here we provide additional evidence to show that the observed saturation is not due to artifacts. We have shown in Figure S2 that the measurements are taken in the linear regime of the $I(V)$ so that current heating-induced saturation can be ruled out. We have also verified that the parasitic resistance of the measurement setup is roughly 200 GΩ, which cannot account for the observed saturation. Finally, we plot in figure S3 $R_{23}^{CNP}(D)$ at $T$ =1.5 K, 10 K and 20 K. All saturate around $D$ = 3.5 V/nm even though $R_{23}^{CNP}$ itself varies by several orders of magnitude. This rules out RF radiation heating as the origin of the observed saturation since such effect should be more pronounced in high-resistance and /or low cooling power (i.e. low temperature) regimes.

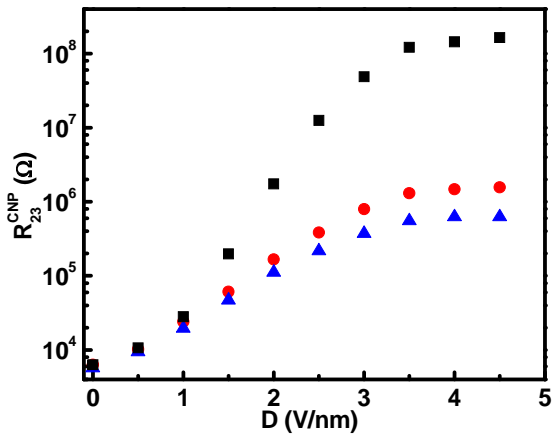

Figure S3. $R_{23}^{\mathrm{CNP}}(D)$ at $T = 1.5$K (black squares), $T = 10$ K (red circles) and 20 K (blue triangles) for positive $D$. All saturate around 3.5 V/nm despite the variation of the resistance value over more than two orders of magnitude.

Below we provide further evidence supporting the saturation of $E_g$ by noting that not only $R_{23}^{CNP}$, but also a few characteristic energy scales related to $E_g$ saturate with increasing $D$. These are extracted from the temperature dependence of $R_{23}^{CNP}$. To ensure the measured $R_{23}^{CNP}(T)$ accurately reflects the $T$-dependence of the dual-gated ABC-TG portion, only $R_{23}^{CNP}(T)$ values greater than 50 kΩ are used. This constrains the temperature and electric field range of useful data. As an example, figure S4a plots $R_{23}^{CNP}(T)$ at $D$ = 4.5 V/nm, where $R_{23}^{CNP}(T)$ increase from ~ 100 kΩ at 200 K to 200 MΩ at 1.5 K.

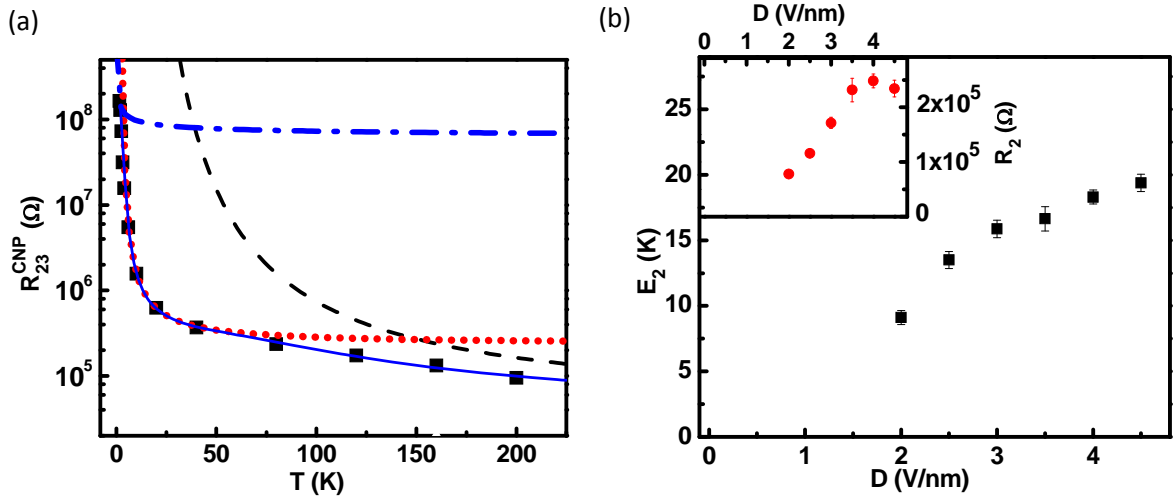

Figure S4. (a) $R_{23}^{CNP}(T)$ at $D$ = 4.5 V/nm (black squares). Black dashed, red dotted, blue dash-dotted and blue solid lines are respectively fittings to the $E_1$, $E_2$ and $T_3$ terms and the sum of all three in eq S1. The fitting parameters are $R_1$ = 36 kΩ, $E_1$ = 301 K, $R_2$ = 2.3x10$^5$ Ω, $E_2$ = 19.4 K, $R_3$ = 5.7x10$^7$ Ω and $T_3$ = 1.57 K. (b) The fitting results of $E_2(D)$ and $R_2(D)$ (inset). For $D$ < 2 V/nm, the contribution of the ABA-TG to $R_{23}^{CNP}(T)$ is non-negligible and data in this regime is not analyzed.

Previous studies[2, 3] in bilayer graphene show that due to disorder induced localized states in the band gap, the *T*-dependence of the CNP resistance cannot be fit to a single activation term but consists of three conduction mechanisms (eq 1 of ref 2):

$$R_{23}^{CNP}(T)^{-1} = R_1^{-1} \exp[-E_1/k_BT] + R_2^{-1} \exp[-E_2/k_BT] + R_3^{-1} \exp[-(T_3/T)^{1/3}] \quad \text{(b)} \quad \text{(S1)}$$

where the $E_1$, $E_2$ and $T_3$ terms represent thermal activation to the mobility edge, nearest neighbor hopping and variable-range hopping respectively.[2] Trilayer devices studied here are similar in fabrication methods and quality to those used in ref 2 and the above mechanisms should also apply. Indeed, data in Figure S4a can be well described by eq S1. Figure S4a plots the three fitting terms separately, together with the total fit. As observed in ref 2, $E_1$ approaches $E_g$ at large *D*. However, our analysis indicates that an unambiguous determination of $E_1$ and $T_3$ is not possible here due to limited temperature range. Therefore the magnitude of $E_g$ is not accessible in our experiment. The uncertainty in $E_1$ and $T_3$, however, has only a weak effect on the fitting of the $E_2$ term, which is dominated by data in the temperature range of 5 - 100 K. The resulting $E_2$ and $R_2$ are plotted in Figure S4b as a function of *D*.

Both $E_2$ (*D*) and $R_2$ (*D*) increase with increasing *D* initially but the rates slow down or saturate at large *D*, similar to the trend displayed by $R_{23}^{CNP}$ and $E_g$ in Figure 2b of the main text. From the theoretical viewpoint, this is perhaps not surprising. The increase of $E_g$ results in smaller density of states of the localized states inside the band gap, which are also more localized in space[4-6]. Consequently, carrier hopping between the localized states becomes more difficult. This explains the experimental trend $E_2$ (*D*) and $R_2$ (*D*) display at small *D*. Following the same argument, the saturation of $E_g$ (*D*) at large *D* stabilizes the characteristics of the localized states and hence the saturation of $E_2$ (*D*) and $R_2$ (*D*).

In summary, although we are not able to determine $E_g$ directly from our data, the tendency toward saturation simultaneously displayed by $E_2$, $R_2$ and $R_{23}^{CNP}$ supports the saturation of the band gap in ABC-TG at large electric field.

1.      Zou, K.; Hong, X.; Keefer, D.; Zhu, J. *Physical Review Letters* **2010,** 105, (12), 126601.

2. Zou, K.; Zhu, J. *Physical Review B* **2010,** 82, (8), 081407.
3. Taychatanapat, T.; Jarillo-Herrero, P. *Physical Review Letters* **2010,** 105, (16), 166601.
4. Nilsson, J.; Castro Neto, A. H. *Physical Review Letters* **2007,** 98, (12), 126801.
5. Rossi, E.; Das Sarma, S. *Physical Review Letters* **2011,** 107, (15), 155502.
6. Li, Q.; Hwang, E. H.; Das Sarma, S. *Physical Review B* **2011,** 84, (11), 115442.